\documentclass{article}
\usepackage{frascatiphys}

\begin{document}

\title{On the Case for a Super $\tau$-Charm Factory
}

\author{
David M. Asner \\
{\em Carleton University, 1125 Colonel By Drive, Ottawa, Ontario, Canada, K1S 5B6}
}
\maketitle
\baselineskip=11.6pt
\begin{abstract}
Design studies for a Super Flavor Factory (SFF), 
an asymmetric energy $e^+e^-$ collider utilizing International Linear
Collider (ILC) techniques and technology, are in progress.
The capablity to run at $\sqrt{s}=3.770$~GeV could be included in the 
initial design.
This report discusses the physics that can be probed with luminosity of 
$10^{35}\,{\rm cm}^{-2} {\rm s}^{-1}$ near $\tau$-charm threshold.
\end{abstract}
\baselineskip=14pt
%par 0: Context:
%The physics programs of the two existing $B$ factories, BABAR and Belle, 
%includes precision studies of CKM physics and $CP$ violation.
%The physics reach of these facilities in many cases is limited by our
%understanding of the strong interaction. Data from charm threshold facilities,
%CLEO-c and BESIII, is essential to benchmark lattice QCD and fully exploit
%the large $B$ factory data sample, in spite of the larger charm sample
%available at the $B$-factories. 

\section{Introduction}
Design studies for a Super Flavor Factory (SFF), 
an asymmetric energy $e^+e^-$ collider utilizing International Linear
Collider (ILC) techniques and technology, are in progress\cite{infnroadmap}.
Energy flexibility is an important component of the design. 
Luminosities of $10^{36}\,{\rm cm}^{-2} {\rm s}^{-1}$ at the $\Upsilon(4S)$ and
$10^{35}\,{\rm cm}^{-2} {\rm s}^{-1}$ at the $\psi(3770)$ are expected.
This report summarizes the physics that can be probed at a 
$\tau$-charm threshold.
The physics case for a Super $\tau$-charm
Factory ($10^{35}\,{\rm cm}^{-2} {\rm s}^{-1}$)
or equivalently the case for designing the SFF with the capability to run at energies near $D\overline D$ threshold must be evaluated relative to the physics
reach with the enormous $\tau$ and charm samples available
from $\sqrt{s}=10.58$~GeV running as well as
 anticipated data samples from CLEO-c\cite{Briere:2001rn} and BESIII\cite{BESIII}. The physics to be probed generally falls into two cateogories: (1) Probes
of QCD to enhance or validate the theoretical control over QCD. This program
will likely be completed with CLEO-c and BESIII (except for charm baryon 
studies). (2) Searches for physics beyond the Standard Model. This program
will not be completed by CLEO-c and BESIII.

%par 1: Background on Bfac

The two existing asymmetric $B$-factories, BABAR at PEP-II and Belle at KEKB,
have operated since 1999. PEP-II is running with peak luminosity of 
$10.0\times 10^{33}\,{\rm cm}^{-2} {\rm s}^{-1}$ and KEKB with peak luminosity
of $15.8\times 10^{33}\,{\rm cm}^{-2} {\rm s}^{-1}$ both at the $\Upsilon$(4S).
The total integrated luminosity recorded by BABAR, 330~fb$^{-1}$, and Belle,
 500~fb$^{-1}$, includes nearly $10^9$ $B\bar B$ pairs. 
Also included in this data sample are nearly $10^9$ $\tau$-pairs and
$4 \times 10^9$ charm mesons. 

%par 2: Complementary facilities, CLEO-c, BESIII

CLEO-c at CESR has been accumulating data in the charm threshold region
since 2003.
CESR is running with a peak luminosity of $7\times 10^{31}\,{\rm cm}^{-2} {\rm s}^{-1}$. Data samples of 30 million $\psi(2S)$, 750~pb$^{-1}$ at $\psi(3770)$,
and 750~pb$^{-1}$ at $\sqrt{s}=4170$~MeV 
are anticipated before shutdown in 2008.
This corresponds to a sample of 2.7 million $D^0 \overline D{^0}$ pairs and 
2.1 million $D^+D^-$ from $\psi(3770)$. The higher energy data sample includes a
comparable number of $D$ mesons plus 700,000 $D_s^{*+}D_s^- + D_s^{+}D_s^{*-}$.
BESIII at BEPCII expects first $e^+e^-$ collisions at the end of 2007 and
design luminosity, $0.6\times 10^{33}\,{\rm cm}^{-2} {\rm s}^{-1}$ at the $J/\psi$, by the end of 2008. Large charmonium samples, $10\times 10^{9}$ $J/\psi$ and  $3\times 10^{9}$ $\psi(2S)$, are anticipated, however the $\tau$ and charm
meson samples will still be much smaller than those from existing $B$-factories.

%par 3: Intro to concept of varying energy

The physics output of a Super Flavor Factory {\it may} 
be maximized by including
energy flexibility in the design of both the accelerator and the detector.
Assuming the facility is optimized for asymmetric collisions at $\Upsilon$(4S),
little performance degradation is expected for center-of-mass energies ranging
from $\Upsilon$(1S) to $\Upsilon$(5S). Including in the initial design the 
flexibility to run at energies in the $\tau$-charm threshold region requires
further study. Some open questions include: 
(1) Should low energy running
be symmetric or asymmetric?
(2) Is the low energy physics reach compromised
by a detector optimized for high energy running?
(3) Is there a physics need for the low energy data given the enormous $\tau$
and charm samples produces at $\sqrt{s}=10.58$~GeV?

\section{The Advantages of Threshold Production}

The production rate of charm during threshold running at a Super Flavor Factory
and $\Upsilon$(4S) running is comparable. Although
the luminosity for charm threshold running is expected to be an order of
magnitude lower, the production cross section is 3 times higher
than at $\sqrt{s}=10.58$~GeV.
Charm threshold data has distinct and powerful advantages over continuum
and $b \to c$ charm production data accumulated above $B$ production threshold.

%\subsection{Charm Events at Threshold are Extremely Clean}
\noindent{\bf Charm Events at Threshold are Extremely Clean:}
The charged and neutral multiplicites in $\psi(3770)$ events are only 5.0 and
2.4 - approximately 1/2 the multiplicity of continuum charm production at $\sqrt{s}=10.58$~GeV. 
%Additionally, $\psi(3770)$ decays are spherical distributing
%decay products uniformly in solid angle. This translates to high efficiency
%and low systematic error.

\noindent{\bf Charm Events at Threshold are pure $D\overline D$:}
No additional fragmentation particles are produced.  The same is true for
$\sqrt{s}=4170$~MeV production of $D\overline D^*$, $D^+_sD^-_s$, and 
$D^+_sD^{*-}_s$.
%, and for threshold produtions of $\Lambda_c \bar \Lambda_c$.
This allows use of kinematic constraints, such as total candidate energy
and beam constrained mass, and permits effective use of missing mass methods
and neutrino reconstruction. The crisp definition of the initial state is a
uniquely powerful advantage of threshold production that is absent in
continuum charm prodution.

\noindent{\bf Double Tag Studies are Pristine:}
The pure production of $D\overline D$ states, together with low multiplicity and
large branching ratios characteristic of many $D$ decays permits effective
use of double-tag studies in which one $D$ meson is fully reconstructed and the
rest of the event is examined without bias but with substantial kinematic 
knowledge. The techniques pioneered by Mark III and extended by 
CLEO-c\cite{He:2005bs} allow
precise absolute branching fraction determination. Backgrounds under these
conditions are heavily suppressed which minimizes both statistical errors and
systematic uncertainties.

\noindent{\bf Signal/Background is Optimum at Threshold:}
The cross section for the signal $\psi(3770) \to D\overline D$ is 
about 1/2 the 
cross section for the underlying continuum $e^+e^- \to$ hadrons background.
By contrast, for $c\bar c$ production at $\sqrt{s}=10.58$~GeV the signal is
only 1/4 of the total hadronic cross section. 
%In addition, the $c\bar c$ fragmentation distributes the final states among 
%many charm hadron species at higher energies.

\noindent{\bf Neutrino Reconstruction:}
%Neutrinos are reconstructed by interpreting the undetected energy and momentum
%as the neutrino four-vector. 
%Neutrino reconstruction is clean. 
The undetected energy and momentum is interpreted as the neutrino four-vector.
For leptonic
and semileptonic charm 
decays the signal is observed in missing mass squared distributions
and for double-tagged events these measurements have low backgrounds. The
missing mass resolution is about one pion mass. For semileptonic decays
the $q^2$ resolution is excellent, about 3 times better than in continuum
charm reconstruction at $\sqrt{s}=10.58$~GeV. Neutrino reconstruction at threshold
is clean.
%Background is negligible at threshold but signficant at $\sqrt{s}=10.58$~GeV.

\noindent{\bf Quantum Coherence:}
%For $D$ mixing and some $CP$ violation studies
The production
of $D$ and $\overline D$ in a coherent $C=-1$ state from $\psi(3770)$ decay is 
of central importance for the subsequent evolution and decay of these 
particles. The same is true for $D\overline D (n)\pi^0 (m) \gamma$ produced
at $\sqrt{s}\sim 4$~GeV where $C=-1$ for even $m$ and $C=+1$ for odd
$m$. The coherence of the two initial state $D$ mesons allows both simple
and sophisticated methods to measure $D\overline D$ mixing parameters, strong
phases, $CP$ eigenstate branching fractions, and $CP$ 
violation\cite{Bigi:1989ah, Asner:2005wf, Gronau:2001nr, Bigi:2000yz, Bianco:2003vb}.

%par 4: Itemize Physics topics
\section{Physics Reach of Low Energy Running}

The low energy data at a Super Flavor Factory can be used to improve our knowledge of QCD, the Standard Model (SM), and search for New Physics (NP) with the 
study of the $\tau$
lepton, charmonia, total hadronic cross section, charm mesons, and charm baryons.
%Current issues in each area are discussed below. 

\subsection{$\tau$ Studies}
%par 5: Physics case for tau threshold running - tau mass and BF are good

Studies of the $\tau$ lepton include measurement of absolute branching 
fractions, precision determination of $\tau$ 
properties - mass, lifetime, and dipole moment - to probe $CPT$ and $CP$
violation, lepton universality, measurements of hadronic currents - 
to probe QCD, measure $V_{us}$ and $m_s$, searches for $CP$ violation in weak
$\tau$ decays,
and searches for rare or Standard Model forbidden decays such as lepton
flavor violation\cite{mroney-dif06}. 
The $\tau$ mass determination benefits from the constrained kinematics of
threshold production but will likely be limited by knowledge of the beam
energy to $\pm 0.1$~MeV at BESIII. This precision can be compared with the
statistical error expected in 100~ab$^{-1}$ of $\pm 0.023$~MeV.
Improved constraints on the $\tau$ neutrino mass in the range of 1-5~MeV
might be attainable at threshold. 
The $\tau$ lifetime cannot be measured in symmetric collisions at threshold. 
The choice of energy for the rest of the $\tau$ program is driven by the 
need for large statistics.
The $\tau$-pair production cross section is 0.89 nb at $\Upsilon$(4S) and
1.2 nb at threshold. Combined with the expected luminosities there is no
compelling reason to run at $\tau$ threshold.

%par 11: Charmonium J/\psi - 4260  see CBX 05-54

\subsection{Charmonium Physics at the $J/\psi$ Resonance}

The design luminosity at a Super Flavor Factory at the $J/\psi$ resonance could
exceed the design luminosity at BESIII by an order of magnitude. The BESIII samples
will be orders of magnitude larger than current BES samples. It is challenging
to predict quantitatively the sensitivities that can be achieved with the 
$10^{12}$ $J/\psi$ sample that might be accumulated at the Super Flavor Factory.
Below are described some of the physics topics that can be elucidated.

\noindent{\bf $J/\psi\to$VS Transitions:}
%\subsubsection{Radiative $J/\psi$ and $\psi(2S)$ Transitions}
%Two radiative processes are to be distinguished: The radiative transition,
%$c \bar c \to \gamma c \bar c$, giving access to the $\eta_c$, and the 
%radiative decay, $c\bar c \to \gamma gg$, giving access to non-charm hadrons
%including glueballs - t
The primary motivation for acquiring a large $J/\psi$ sample is 
the study of glueballs. More generally, it is the study of scalar ($J^{PC}\!=\!0^{++}$) and
tensor ($J^{PC}\!=\!2^{++}$) mesons produced in radiative $J/\psi$ decay.
The Particle Data Group lists three light isoscalar mesons, the $f_0(1370)$, 
$f_0(1500)$, and $f_0(1710)$ while the quark model predicts two in this mass
region. There is a strong bias that one of the $f_0$ is {\it the} glueball
or that
the three observed $f_0$ mesons are mixtures of the glueball and the two $q\bar q$ states. The
double radiative decay $J/\psi \to \gamma f_0$ followed by
$f_0 \to \gamma \rho, \gamma \phi$\cite{Close:2001ga,Close:2005vf}
and the decay $J/\psi \to V f_0$\cite{Close:2005vf,Ablikim:2004wn,Ablikim:2004st,Zhao:2005ip}
may elucidate the glueball content of the $f_0$ mesons.

\noindent{\bf Exclusive ${\bf J/\psi}$ Branching Fractions:}
%\subsubsection{${\bf J/\psi}$ and $\psi(2S)$ Branching Fractions}
Models of charmonium make definite predictions for the $\psi(2S)$ branching fractions based on
$J/\psi$ decay rates. Exclusive branching fractions are not well measured.
A systematic program of precision $J/\psi$ 
branching ratio measurements is desirable to test and improve our understanding
of charmonium. Hadronic decays to charmless final states account for 87\% of $J/\psi$ decay - only 40\% of that is measured exclusively.

\noindent{\bf Leptonic $J/\psi$ Decays:}
%\subsubsection{Leptonic $J/\psi$ Decays}
The $J/\psi$ is commonly identified through its decay to leptons. 
The relative branching ratios of $J/\psi$ to leptons are known to 1\% and
do not pose a precision barrier to other measurements. The only exception is
in the determination of $\Gamma_{ee}(J/\psi)$ through ISR\cite{Aubert:2003sv}.

\noindent{\bf Threshold Enhancements:}
%\subsubsection{Threshold Enhancements}
BES reported an enhancement in the $M(p\bar p)$ spectrum in $J/\psi \to 
\gamma p \bar p$ decays which they attribute to a sub-threshold resonance,
the mass (1859 MeV) and width ($\Gamma <$ 30 MeV at 90\% C.L.) of which
are not consistent with any known particle hence it is called the $X(1860)$\cite{Bai:2003sw}. 
An enhancement observed by BES in the $\pi^+\pi^-\eta^\prime$ spectrum in 
$J/\psi \to \gamma\pi^+\pi^-\eta^\prime$ may confirm this observation\cite{Ablikim:2005um}. 
However,
the enhancement is not observed 
in $J/\psi \to \pi^0 p \bar p$ decays at BES or
in $\Upsilon$(1S) $\to \gamma p \bar p$ decay at CLEO\cite{Athar:2005nu}. 
BES has observed additional threshold enhancements\cite{Ablikim:2006dw,unknown:2006ca}.
BESIII will accumulate a significant $J/\psi$ data sample which may
clarify the situation.
%{\bf Conclusion: $J/\psi$ Running}
%\subsubsection{Conclusion: $J/\psi$ Running}

\subsection{Charmonium Physics at the $\psi(2S)$ Resonance}
The design luminosity at a Super Flavor Factory at the $\psi(2S)$ resonance could
exceed the design luminosity at BESIII by an order of magnitude. The BESIII samples
will be orders of magnitude larger than current BES (58 million) and anticipated CLEO-c (30 million) 
samples. It is challenging
to predict the physics issues to be addressed with $10^{12}$ $\psi(2S)$ sample accumulated at the SFF.

\noindent{\bf Radiative $\psi(2S)$ Transitions:}
The $\eta_c$, the charmonium ground 
state, is of particular interest since its bottomonium counterpart is
unobserved experimentally. The $\eta_c$ mass and width have been determined
utilizing multiple production mechanisms but without good agreement. 
Additionally, the exclusive decay modes measured account for about 25\%
of the total with substantial uncertainties. Relating the $\eta_c$ branching
fractions to the corresponding measurements on the $\eta^\prime_c$
tests our models of charmonium. Although $\eta_c$ production rate is greater from
radiative $J/\psi$ decay, data accumulated at the $\psi(2S)$
allows for both 
$\psi(2S) \to \gamma \eta_c$ and $\psi(2S)\to \gamma \eta^\prime_c$ processes.

\noindent{\bf Exclusive $\psi(2S)$ Branching Fractions:}
Models of charmonium make definite predictions for the $\psi(2S)$ based on
$J/\psi$ decay rates. Exclusive branching fractions are not well measured.
A systematic program of precision $\psi(2S)$ 
branching ratio measurements at CLEO-c and BESIII will test our understanding of charmonium.

\noindent{\bf Properties of $h_c(1^1P_1)$:}
%\subsubsection{Properties of $h_c(1^1P_1)$}
In 2005, CLEO reported first observation of the $h_c$\cite{Rosner:2005ry,Rubin:2005px}. 
%decaying by the radiative
%transition $h_c \to \gamma \eta_c$. 
The $h_c$ is the last of the eight
bound states of charmonium expected to lie beneath open charm threshold.
Precision determination of the properties of the $h_c$, mass, width, branching
fractions will improve our understanding of the QCD potential -- in particular,
measuring $M(h_c)$ relative to the center of gravity of the $\chi_{cj}$ states. 
Predictions using a number of theoretical models\cite{Appelquist:1978aq,Godfrey:2002rp} span a wide range of
values and therefore precise measurements of the hyperfine splitting can 
distinguish among models.
A precision of better than 0.1 MeV on $M(h_c)$ and 0.2 MeV on $\Gamma(h_c)$ is
expected from a sample of 300 million $\psi(2S)$ - 10\% of the sample anticipated
in one year at BESIII.

\noindent{\bf $J/\psi$ Sample from $\psi(2S)$:}
%\subsubsection{$J/\psi$ Sample from $\psi(2S)$}
A large $J/\psi$ sample can be obtained through the cascade production
from $\psi(2S)$ decays to $\pi^+\pi^- J/\psi$ with ${\cal B}=33.5\%$.
This reduces and perhaps eliminates the need for running at the $J/\psi$.

\noindent{\bf $\chi_{cj}$ Sample from $\psi(2S)$:}
%\subsubsection{$\chi_{cj}$ Sample from $\psi(2S)$}
The $\psi(2S)$ is a factory for producing $\chi_{cj}$ through $\psi(2S) \to
\gamma \chi_{cj}$. The branching ratios are around 9\% and photon detection
efficiencies (after $\pi^0$ suppression) are about 50\% so that one detectable
each of $\chi_{c0}$, $\chi_{c1}$ and $\chi_{c2}$ are produced for every 20 
$\psi(2S)$. Hadronic decay of the $\chi_{cj}$ offer a number of interesting
measurements (1) study of mesons with manifestly exotic quantum numbers through
the Dalitz plot analysis of $\chi_{cj} \to \eta \pi\pi$ (2) study of 
the quark content of the three isoscalar mesons,
$f_0(1370)$, $f_0(1500)$, and $f_0(1710)$. Since $f_2(1270)$
is nearly entirely $u \bar u, d\bar d$ 
and the $f^\prime_2(1525)$ is nearly entirely
$s\bar s$, these two narrow resonances can be used to tag the flavor
of the $f_0$. Therefore comparing the relative rates of $\chi_{c2} \to f_2(1270)f_0$ and $\chi_{c2} \to f^\prime_2(1525)f_0$ 
can shed light on the quark content
of the $f_0$ in question. (3) Multibody charmonium decays access most of the 
kinematic range accessible to $B$ decays. 
Improved modeling of $\pi\pi$, $K\pi$, $KK$ $S$-wave in charmonium samples,
such as $\chi_{c0}\to \pi^+\pi^-K^+K^-$\cite{Ablikim:2005kp}, will
enable precision measurements of CKM angles $\alpha$, $\beta$, and $\gamma$.
Specifically, understanding the $\pi\pi$ S-wave is important for the determination of $\alpha$ 
($B^0 \to \pi^+\pi^-\pi^0$) and $\gamma$ ($B^- \to D K^-, D \to K^0_S\pi^+\pi^-$),
understanding the $K\pi$ S-wave is important for the detemination of $\gamma$ 
 ($B^- \to D K^-, D \to K^0_SK\pi$), and the understanding of the $KK$ S-wave is important
of the detemination of $\beta$ using $b\to s$ penguin processes ($B \to K^0_S K^+K^-$).

\noindent{\bf The $M2/E1$ Ratio for $\chi_{cj} \to \gamma J/\psi$:}
The transitions $\chi_{c1} \to \gamma J/\psi$ and $\chi_{c2} \to \gamma J/\psi$
can each proceed through either an $E1$ or $M2$ transition. Current measurements are at odds with theoretical expectations\cite{Armstrong:1993fk,Ambrogiani:2001jw}. The BESIII sample of $\psi(2S)$
will be sufficient to conclusively measure this multipole ratio.

%\noindent{$\eta_c$ and $\eta_c^\prime$ Sample from $\psi(2S)$:}

\subsubsection{Using $\psi(2S)$ Running as Calibration Data}
Data taken at the $\psi(2S)$ has a utilitarian value. The processes
$\psi(2S) \to \pi^+\pi^- J/\psi$, $\pi^0\pi^0 J/\psi$, and $\gamma\chi_{cj}$ 
are particularly useful. They provide hadron and photon spectra of known 
momenta that are useful for measuring reconstruction efficiencies in data, 
Monte Carlo tuning, and detector calibration. Such data, taken at intervals
interspersed between higher energy running, can be used for monitoring
software and detector performance, leading to better control of systematic
uncertainties for studies of $\tau$ and charm at threshold.

%\noindent{Conclusion: $\psi(2S)$ running}

\subsection{Charmonium Physics above $c\bar c$ Threshold}

\noindent{\bf The total $e^+e^-$ cross section and $R$ measurement:}
%\subsubsection{The total $e^+e^-$ cross section and $R$ measurement}
The BES 
measurements of the total cross section of $e^+e^-$ annihilation to hadrons
or $R$ scan from 85 energy points between 2 and 5~GeV have an average
statistical error of 6.6\% and a systematic uncertainty of 3.3\% using about 
70~nb$^{-1}$ per point\cite{Bai:2001ct}.
The CLEO measurement of $R$ with an accuracy of 2\% in the 
vicinity of the $\Upsilon$(4S)\cite{Ammar:1997sk} and recent progress in the calculation of
radiative corrections together with the hermeticity of the BESIII detector
allows one to expect a systematic uncertainty of about 1\% below and 2-4\% above
open charm threshold.

A second approach to measuring $R$ in the 2-5~GeV range uses initial state 
radiation (ISR) from 10.58~GeV\cite{Anulli:2004nm}. This approach should be competitive with results
from the $R$ scan. Existing data samples from BABAR and Belle are expected to
be limited by systematic uncertainties to a precision of a few percent.
If further advances in the calculation of radiative corrections enable $R$
measurements with sub-percent precision the ISR data from the $\Upsilon$(4S) 
running of a SFF will have more than sufficient statistics
for the $R$ measurement. 
%Thus, there is no compelling reason to do an $R$ scan
%at the Super Flavor Factory.

\noindent{\bf $c\bar c$ Hybrids, $Y(4260)$ and Other Resonances:}
%\subsubsection{$Y(4260)$ and Other Resonances}
The region at center-of-mass energies above open charm production
threshold is of great interest to theory due to its richness of $c \bar c$ 
states, the properties of which are not well-understood. Prominent structures
in the hadronic cross section are the $\psi(3770)$, $\psi(4040)$, and 
$\psi(4160)$. 
% 50 points, 7 1/pb at CLEO-c = 300 days
% 50         70     at BESIII = 300 days
% 50         7 1/fb at SFF = 300 days 
% 100         3.5 1/fb at SFF = 300 days 
% 100         1 1/fb at SFF = 100 days 
A dedicated scan, 1~fb$^{-1}$/per energy point at 100 energy points, to search
for hybrid $c \bar c$ mesons coupling directly to $e^+e^-$ through
a virtual photon or produced in decay products of charmonia represented by the
structure in the hadronic cross section would require 100 days of SFF running.

Recently, observations of new charmonium-like states decaying
to open-charm have been reported in this energy region. Additionally, an
enhancement, the $Y(4260)$, in the invariant mass spectrum of $\pi^+\pi^- J/\psi$ has been
observed by BABAR in ISR\cite{Aubert:2005rm} and in $B \to K(\pi^+\pi^- J/\psi)$\cite{Aubert:2005zh}. 
This observation
was confirmed by CLEO is both ISR ($\pi^+\pi^- J/\psi$)\cite{cleocISR} and in $e^+e^-$ 
collision at $\sqrt{s}=4260$~MeV ($\pi^+\pi^- J/\psi$, $\pi^0\pi^0 J/\psi$, $K^+K^- J/\psi$)\cite{Coan:2006rv}. These observation help distinguish among the many models 
predicting properties of the $Y(4260)$.

%\subsubsection{Conclusion}

%\subsubsection{Charmonium at a $\sqrt{s}=10.58$~GeV}

%par 7: charm baryons
\subsection{$D^0$, $D^+$, and $D^+_s$ Decays}

\noindent{\bf Leptonic Charm Decays - $D^+ \to \ell^+\nu$, $D^+_s \to \ell^+\nu$:}
For the muonic decays, CLEO-c and BESIII, will determine the decay constants $f_D$ and $f_{D_s}$,
to precision of 1\%. The decay constants measure the nonperturbative wave function of the meson at
zero inter-quark separation. In the Standard Model the relative widths for $\tau^+\nu$, $\mu^+\nu$ and $e^+\nu$ are
2.65 : 1: 2.3$\times 10^{-5}$. Comparison of electronic, muonic and tauonic rates\cite{Bonvicini:2004gv,Artuso:2005ym,Rubin:2006nt} 
tests for physics beyond the SM.

\noindent{\bf Exclusive Semileptonic Charm Decays - $D\to (K,K^*,K\pi)\ell\nu$, $D\to (\pi,\eta,\rho,\omega,\pi\pi)\ell\nu$, $D_s \to (\eta, \phi) \ell \nu$, $D_s \to (K,K^*,K\pi)\ell\nu$, and $\Lambda_c \to \Lambda \ell \nu$:}
Absolute branching ratios, for the $D$ and $D_s$, will be measured to $\sim 1\%$ and the form factor slopes to $\sim 2\%$ by 
CLEO-c\cite{Coan:2005iu,Huang:2005iv,Artuso:2005jw}
and BESIII. Threshold production enables form factor measurements with improved resolution over the full range of 
$q^2$. Semileptonic decay rates {\it and} accurate knowledge of form factors are required for CKM elements $|V_{ub}|$, $|V_{cb}|$, $|V_{cd}|$, and $|V_{cs}|$.

\noindent{\bf Inclusive Semileptonic Charm Decays - $D \to \ell X$, $D_s \to \ell X$, and $\Lambda_c \to \ell X$:}
Inclusive branching ratios, for the $D$ and $D_s$, will be measured to $\sim 1\%$ by
CLEO-c\cite{Adam:2006nu} and BESIII.
Inclusive spectra have three advantages compared to semileptonic branching fractions
as a probe of theory: (1) Theoretical interpretation is cleaner as spectra are
independent of the hadronic width. (2) Spectra contain both shape and rate 
information. (3) Non-perturbative effects are pronounced in the lepton endpoint 
region. 
%HQET provides relations among inclusive semileptonic charm decays and betweenincluse charm and inclusive $B$ processes. 
Low backgrounds associated with tagged events at threshold enable inclusive 
semileptonic studies. The theoretical description of semileptonic decays in the 
charm sector, coupled with heavy quark symmetry, will improve the description of 
semileptonic $B$ decays and the determination of $V_{ub}$ and $V_{cb}$.

\noindent{\bf Hadronic Charm Decays:}
CLEO-c and BESIII will measure the branching fractions for the normalizing modes $D^0 \to K\pi$, 
$D^+\to K^-\pi^+\pi^+$, and $D^+_s \to K^+K^-\pi^+$ to a precision of less than 1\%\cite{He:2005bs,Li:2006nv}. 
Sensitivity to
Cabibbo suppressed modes, $D \to (n)\pi (m)\pi^0$, at $10^{-5} - 10^{-6}$ level\cite{Rubin:2005py,Li:2006nv} and independent measurements
of $D \to K^0_S X$ and $D \to K^0_L X$ are possible at threshold due to low backgrounds and constrained kinematics
with these data samples.

\subsection{Impact on CKM physics}
%par 8: detail the import of CKM magnitudes at CLEO-c/BESIII

\noindent{\bf Determination of $V_{ub}$:}
Limited by form factor calculations
to $\sim$ 13\%. Improving form factor calculation methods in the charm decays 
$D\to \pi \ell \nu$ and $D\to \rho \ell \nu$ with data from CLEO-c
will enable 5\% precision in $\left|V_{ub}\right|$. 
Improved understanding of weak annihilation contributions to inclusive semileptonic
charm decay will improve the extraction of $V_{ub}$ from
inclusive semileptonic $B$ meson decay.
%Additional i
Improvement
will be possible with the BESIII data sample.

\noindent{\bf Determination of $V_{td}$ and $V_{ts}$:}
Limited by ignorance of $f_B\sqrt{B_{B_d}}$ and $f_{B_s}\sqrt{B_{B_s}}$.
Determining $\left| V_{td}\right|$ and $\left| V_{ts}\right|$ from $B$ mixing
measurements requires improved determination of $f_B$ and $f_{B_s}$.
Precision measurements of $f_D$, $f_{D_s}$ and $f_D/f_{D_s}$ at CLEO-c and BESIII
will enable the necessary theoretical advances. 
%{\bf Say something about Belle B to tau nu}

\noindent{\bf Determination of $V_{cd}$ and $V_{cs}$:}
Currently known to $\sim$10\% level by direct measurement. CLEO-c is measuring
absolute branching ratios of leptonic and semileptonic decays from which 
$\left| V_{td}\right|$ and $\left| V_{ts}\right|$ can be determined with few
percent accuracy. 
Again form factor and decay constant calculations must achieve
comparable precision 
%and be validated by the breadth of CLEO-c measurements
for the few percent precision on CKM parameters to be realized.
These measurements will enable Unitarity tests of the CKM matrix.
Data from BESIII will enable additional improvement.

\noindent{\bf Determination of $V_{cb}$:}
%Leptonic charm decays: $D^- \to \ell \bar \nu$ and  $D_s^- \to \ell \bar \nu$
Presently limited by several factors including
theoretical control of form factors and experimental determination of 
${\cal B}(D\! \to\! K\pi)$. CLEO-c will drive form factor technology and will
measure the normalizing hadronic charm branching ratios at the percent level.
Precision of 3\% in $V_{cb}$ is expected. Improved precision using BESIII data will require
theoretical advances to better estimate corrections to form factors. 
The inclusive determination of $V_{cb}$ will benefit from
better knowledge of the inclusive lepton spectra which refine modeling of the 
``cascade'' decays $b\to c\to s\ell \nu$.

%\subsubsection{Multibody nonleptonic charm decays and CKM angles $\alpha$, $\beta$, and $\gamma$}
%par 9: ckm angles

\noindent{\bf CKM Angle $\gamma/\phi_3$:}
Measurement of the CKM angle $\gamma/\phi_3$ is challenging. Several methods
have been proposed using $B^\mp \to D K^\mp$ decays; (1) the 
Gronau-London-Wyler (GLW)\cite{glw} method where the $D$ decays to $CP$ eigenstates
(2) the Atwood-Dunietz-Soni (ADS)\cite{ads} method where the $D$ decays to flavor eigenstates
and (3) the Dalitz-plot method\cite{Giri:2003ty} where the $D$ decays to a three-body final state.
Uncertainties due to charm contribute to each of these methods.
%The charm physics program includes a 
A variety of charm measurements impact
the determination of $\gamma/\phi_3$ from the $\Upsilon$(4S):
%The pertinent components of this program are 
(1) Improved constraints on charm 
mixing amplitudes - important for GLW, (2) Measurement of the relative rate and 
relative strong phase $\delta$ between
$D^0$ and $\overline D{^0}$ decay to $K^+\pi^-$ - important for ADS, and 
(3) studies of charm Dalitz plots tagged by hadronic flavor or $CP$ eigenstates.
Charm threshold data is necessary for the 
measurement of $\delta$ and the study of $CP$ tagged Dalitz plots. Only 1~fb$^{-1}$ of charm
threshold data is required to measure $\cos\delta$ to $\pm 0.1$\cite{Asner:2005wf} 
which will be accomplished by
CLEO-c and BESIII. The CKM angle $\gamma/\phi_3$ will be measured to $1^o$ (statistical) 
with 100~ab$^{-1}$ of SFF data. The sample of 30~fb$^{-1}$ at charm threshold - expected from
BESIII - is needed to limit Dalitz-plot systematic uncertainties to $1^o$\cite{Bondar:2005ki}.

%par 12: Rare Charm, mixing, CPV
\section{$D$ Mixing, $CP$ Violation, and Rare Charm Processes}

%\section{$D^0\bar{D}^0$-Mixing}
\noindent{\bf ${\bf D^0-\bar{D}^0}$ Mixing:}
Neutral flavor oscillation in the $D$ meson system is highly suppressed
within the SM. The time evolution of a particle produced as a $D^0$
or $\bar{D}^0$, in the limit of $CP$ conservation, is governed by four parameters:
$x=\Delta m/\Gamma$, $y=\Delta \Gamma/2\Gamma$ characterize the mixing matrix, 
$\delta$ the relative strong phase
between Cabibbo favored (CF) and doubly-Cabibbo suppressed (DCS) amplitudes and 
$R_D$ the DCS decay rate relative to the CF decay rate\cite{Asner:2004gi}. 
%Standard Model based predictions for $x$ and $y$, as well as a variety of non-Standard 
%Model expectations, span several orders of magnitude\cite{Petrov:2003un}.
%It is reasonable to assume that $x\approx y \approx 10^{-3}$ in the Standard Model.
The mass and width differences $x$ and $y$ can be measured in a variety of ways.
The most precise limits are obtained by exploiting the time-dependence of 
$D$ decays\cite{Asner:2004gi}. A time-dependent analysis of
$D^0 \to K^0_S\pi^+\pi^-$ Dalitz plot\cite{Asner:2005sz} allows simultaneous determination of
$x$ and $y$ without phase or sign ambiguity but with $\sim 4\times$ less sensistivity relative to the time-dependent study of $D^0 \to K^+\pi^-$\cite{Godang:1999yd}. 
Time-dependent analyses are not feasible at 
CLEO-c and BESIII; however,
the quantum-coherent $D^0\bar{D}^0$ state provides time-integrated sensitivity 
to $x$, $y$ at ${\cal O}(1\%)$ level and $\cos\delta\sim 0.05$\cite{Briere:2001rn,Asner:2005wf}. 
Asymmetric collisions near $\sqrt{s}=4$~GeV at a SFF could enable time-dependent
measurements.
%Although CLEO-c does not have sufficient sensitivity to observe Standard
%Model charm mixing the
%These projected results compare favorably with current experimental results; 
%see Fig.~1 in Ref.\cite{Asner:dq}.

%\section{$CP$ Violation}
\noindent{\bf ${\bf CP}$ Violation\cite{Bigi:1989ah, Bigi:2000yz}:}
Standard Model $CP$ violation is strongly suppressed in charm as the effective
weak phase is rather small - ${\cal O}(\lambda^4)$, arising only in 
singly-Cabibbo-suppressed transitions where expected $CP$ asymmetries reach ${\cal O}(0.1\%)$. Significantly larger values would indicate NP. Any asymmetry in CF or DCS decays requires new physics - except
for $D^\pm \to K^0_S \pi^\pm$, where the $CP$ impurity due to the $K^0_S$
induces an asymmetry of $3.3\times 10^{-3}$.
%Standard Model predictions for the rate of $CP$ violation in charm mesons are as large
%as 0.1\% for $D^0$ decays and as large as 1\% for certain $D^+$ and $D^+_s$ decays\cite{Buccella:1996uy}.
%In addition to indirect $CP$ violation, both SM and NP effects can 
%induce different contributions to the decay amplitudes of $D$ mesons.
%This phenomenon can be traced back to the appearance of complex-valued
%couplings (CKM parameters) in the $\Delta C = 1$ Lagrangian that mediates $D$
%decays and leads to a $CP$-violating difference between decay rates of 
%$CP$-conjugated states.
%\par
%{\bf need two sentences!
At $\sqrt{s}=10.58$~GeV, the decay channel $D^{*\pm} \to D\pi^\pm$ is used to provide
a flavor tag for the $D$ meson. Threshold production provides a $CP$ tag.
$CP$ conservation forbids certain final states from the decay of the 
$D^0\overline D{^0}$ pair from correlated production. For $C=-1$ states produced
from $\psi(3770) \to D\overline D$, the final state $f_+f_+$, such as $(K^+K^-)(\pi^+\pi^-)$, is
forbidden; For $C=+1$ states produced
from $e^+e^-(4170\,{\rm MeV}) \to D\overline D\gamma$, the final state $f_+f_-$, such as
$(K^+K^-)(K^0_S\pi^0)$, is forbidden.
The expected sensitivity to direct $CP$ violation with tagged decays at CLEO-c and BESIII is $\sim 1\%$; at a SFF it is $\sim 0.1\%$. 
There are also methods to probe $CP$ violation with untagged charm 
decays\cite{Petrov:2004gs}.

%{\noindent \bf Dalitz Plot Analyses:} 
%A Dalitz plot analysis of multibody final states measures amplitudes and phases
%rather than the rates and so may provide greater sensitivity to $CP$ violation.
%In the limit of $CP$ conservation, charge conjugate decays will have the same
%Dalitz distribution. Although the $D^+$ and $D_s^+$ decays are self-tagging,
%there have been no reported Dalitz analyses that search for $CP$ violation with
%charged $D$'s. The decay $D^0 \to K_S \pi^+\pi^-$ proceed through intermediate
%states that are $CP+$ eigenstates, such as $K_S f_0$, $CP-$ such as $K_S\rho$ and flavor eigenstates such as $K^{*-}\pi^+$\cite{Muramatsu:2002jp}. 
%It is noteworthy that for uncorrelated $D^0$
%the interference between $CP+$ and $CP-$ eigenstates integrates to zero across the 
%Dalitz plot but for correlated $D$ the interference between $CP+$ and $CP-$ eigenstates
%is locally zero. 
Decays to final states with more than two pseudoscalars or one pseudoscalar and one
vector meson contain more dynamical information than given by their widths.
Distribution on Dalitz plots or $T$ odd moments can exhibit $CP$ asymmetries
considerably larger than those for the width.
The Dalitz plots for $\psi(3770) \to D^0\bar{D}^0 \to f_+K_S\pi^+\pi^-$ and  $\psi(3770) \to D^0\bar{D}^0 \to f_-K_S\pi^+\pi^-$ will be
distinct
%,as illustrated in Fig.~\ref{fig1}, 
and the Dalitz plot for the untagged sample  $\psi(3770) \to D^0\bar{D}^0 \to X K_S\pi^+\pi^-$ will be distinct from that observed with uncorrelated $D$'s from continuum production at $\sim 10$~GeV\cite{Muramatsu:2002jp}.
The sensitivity at charm threshold to $CP$ violation with Dalitz plot analyses has not yet been evaluated.
The sensitivity to $CP$ violation with flavor tagged $D^0 \to K^0_S\pi^+\pi^-$ at $\sqrt{s}=10.58$~GeV
in 9~fb$^{-1}$ is in the range (3.5 to 28.4)$\times 10^{-3}$ depending on the decay channel\cite{Asner:2003uz}.

%\section{Rare Charm Decays}
\noindent{\bf Rare and Forbidden Charm Decays:}
The Standard Model predicts vanishingly small
branching ratios for processes such as $D \to \pi/K^{(*)} \ell^+\ell^-$
which is GIM suppressed.
Rare decays of charmed mesons and baryons provide ``background-free''
probes of new physics effects that enhance rare or allow forbidden
processes such as lepton flavor violation ($D \to Ke^+\mu^-$)
and lepton number violation ($D^+ \to K^-e^+e^+$). 
Current limits are ${\cal O}(10^{-4})$ to ${\cal O}(10^{-6})$ and are limited by statistics.
Limits from BESIII will improve to ${\cal O}(10^{-7})$ to ${\cal O}(10^{-8})$\cite{Li:2006nv}.
It is noteworthy that long
distance dynamics can generate some of these final states even 
within the SM such as $D^+ \to \phi \pi^+$, $\phi \to e^+e^-, \mu^+\mu^-$. In 
these cases distinguishing NP from SM contributions will be a challenge.

\subsection{Charm Baryons}
The absolute scale for charm baryon decays decays is not well determined. To
date, the measurements of the charm baryon absolute rate are model
dependent. The decay mode used to normalize all other decay rates is
$\Lambda_c^+ \to pK^-\pi^+$ but it is only known to be 
%$3.0\% < {\cal B}(\Lambda_c) < 9.7\%$ at 90\% confidence level. 
$(5.0\pm1.3)\%$\cite{Burchat:2004gx}.
Prospects for improving this situation are poor without data taken
near $\Lambda_c \bar \Lambda_c$ threshold where a modest amount data, 20~fb$^{-1}$, could
measure ${\cal B}(\Lambda_c)$ to within 1\% of itself.
Semileptonic decays of charm baryons are also of interest to test our
theoretical understanding of form factors. 
%Comparing ${\cal B}(D \to X\ell \nu)$ and  ${\cal B}(\Lambda_c \to X\ell \nu)$ provides a test of HQET.
Measuring $\Lambda_c$ is important for understanding of $b$ fragmentation
and the study of $\Lambda_b$ which is usually studied in $\Lambda_b \to \Lambda_c X$, $\Lambda_c \to pK^-\pi^+$.

%par 10: Discuss reach of SuperB factories
%\section{Physics at $\sqrt{s}=10.58$~GeV}

%par 13: Introduce idea of flexible energy machine super flavor factory
%par 14: Discuss detector/accelerator difficulties
%par 15: Choosing a machine to accompany the LHC
%par 16: Skeleton Run plan for SFF
%\section{Skeleton Run Plan for the Super Flavor Factory}
%par 17: summarize requirements - see end of Roney talk.
%\section{Requirements for a Super Flavor Factory}
%par 18: Conclusion: See slides 18-20
\section{Summary}
Current questions in $\tau$, open charm and charmonium have been summarized.
For $\tau$ physics there is no compelling reason to run near threshold.
Charmonium data samples will increase by several orders of magnitude before a
2011 as BESIII carries out it physics program. 
It is difficult to forecast what questions will remain and new question
arise in this area. Although ISR from $\sqrt{s}=10.58$~GeV renders an $R$ scan
unneccessary, the capability to study charm hybrids and 
charmonium above open charm threshold
with direct production is desireable.
Using open charm produced near threshold 
to improve the precision of CKM parameters
determined by BABAR, Belle, CLEO-c and BESIII will require theoretical advances.
Searches for $D$-mixing, $CP$ violation charm and rare charm decays are driven by statistics.
The enormous $B$ and charm samples available at a Super Flavor Factory at $\sqrt{s}=10.58$~GeV
partially mitigate the need for charm threshold data.
However,
there are many advantages to running at threshold such as lower backgrounds, lower multiplicity,
increased kinematical constraints and quantum coherence. Thus, it is prudent to design the Super Flavor
Factory machine and detector with the flexibility to operate effetively with center-of-mass energies
ranging from $J/\psi$ to $\Upsilon$(5S). More work is required and is ongoing to determine 
whether the benefit of charm threshold running is worth the cost and effort.

\end{document}